\documentclass[reprint,superscriptaddress,nofootinbib,amsmath,amssymb,aps,pra]{revtex4-2}

\usepackage{graphicx}
\usepackage{dcolumn}
\usepackage{bm}
\usepackage{braket}
\usepackage{mathrsfs}
\usepackage{hyperref}
\hypersetup{
    colorlinks,%
    citecolor=blue,%
    filecolor=blue,%
    linkcolor=blue,%
   	urlcolor=blue,
   	linktoc=page
}

\begin{document}

\preprint{APS/123-QED}

\title{Carrollian Perspective on Celestial Holography}



\author{Laura Donnay}
 \email{laura.donnay@tuwien.ac.at}
 \affiliation{%
 {{Institute for Theoretical Physics, TU Wien}}\\
{{ Wiedner Hauptstrasse 8–10/136, A-1040 Vienna, Austria}}
}%
\author{Adrien Fiorucci}
 \email{adrien.fiorucci@tuwien.ac.at}
 \affiliation{%
 {{Institute for Theoretical Physics, TU Wien}}\\
{{ Wiedner Hauptstrasse 8–10/136, A-1040 Vienna, Austria}}
}%
\author{Yannick Herfray}
 \email{yannick.herfray@umons.ac.be}
\affiliation{%
 {Service de Physique de l'Univers, Champs et Gravitation,
Universit\'e de Mons} \\ 
{20 place du Parc, 7000 Mons, Belgium}
}

\author{Romain Ruzziconi}
 \email{romain.ruzziconi@tuwien.ac.at}
\affiliation{%
 {{Institute for Theoretical Physics, TU Wien}}\\
{{ Wiedner Hauptstrasse 8–10/136, A-1040 Vienna, Austria}}
}%

\date{\today}

\begin{abstract}
We show that a $3d$ sourced conformal Carrollian field theory has the right kinematic properties to holographically describe gravity in $4d$ asymptotically flat spacetime. The external sources encode the leaks of gravitational radiation at null infinity. The Ward identities of this theory are shown to reproduce those of the $2d$ celestial CFT after relating Carrollian to celestial operators. This suggests a new set of interplays between gravity in asymptotically flat spacetime, sourced conformal Carrollian field theory and celestial CFT.
\end{abstract}

\maketitle

\section{Introduction}
\label{sec:Introduction}
The holographic principle \cite{tHooft:1993dmi,Susskind:1994vu} states  that  gravity in a given spacetime region can be encoded on a lower-dimensional boundary of that region. Extending this paradigm beyond the celebrated Anti-de Sitter (AdS) / Conformal field theory (CFT) correspondence \cite{Maldacena:1997re,Witten:1998qj,Aharony:1999ti} 
to the more realistic model of asymptotically flat spacetimes is part of an intensive ongoing research effort, referred to as \textit{flat space holography} (see \cite{Susskind:1998vk,Polchinski:1999ry,Giddings:1999jq,deBoer:2003vf,Arcioni:2003td,Arcioni:2003xx,Mann:2005yr} for early works). The
asymptotic symmetries preserving the boundary structure of asymptotically flat spacetimes form the Bondi-van der
Burg-Metzner-Sachs (BMS) group \cite{Bondi:1962px,Sachs:1962wk,Sachs:1962zza}, which is an infinite-dimensional enhancement of the Poincar\'e group with
supertranslations. 

Two different roads, referred here as ``Carrollian'' and ``Celestial'' holographies, have emerged in order to describe quantum gravity in $4d$ asymptotically flat spacetime and might seem in apparent tension. 

In the first picture, the dual theory is proposed to be a BMS field theory living on the $3d$ null boundary of the spacetime~\cite{Dappiaggi:2004kv,Dappiaggi:2005ci,Bagchi:2016bcd,Bagchi:2019xfx,Bagchi:2019clu,Laddha:2020kvp,Chen:2021xkw,Bagchi:2022owq} or, equivalently~\cite{Duval:2014uva,Duval:2014lpa}, a conformal Carrollian field theory.
While this $4d$ bulk / $3d$ boundary theory point of view follows the familiar pattern of a codimension-one holographic duality, establishing such a flat space holographic dictionary presents new challenges that one is not used to encounter in AdS/CFT. Key differences include, on the one hand, the null nature of the conformal boundary and, on the other hand, the presence of radiative flux leaking through this boundary. However, this approach is suggested by a flat limit process in the bulk, which consists of taking the cosmological constant to zero. This implies an ultra-relativistic limit on the boundary theory contracting the conformal symmetries into BMS symmetries, see \cite{Barnich:2012aw,Barnich:2012xq,Bagchi:2012xr,Bagchi:2012cy,Detournay:2014fva,Bagchi:2014iea,Bagchi:2015wna,Hartong:2015usd} for successful applications in $3d$ gravity and \cite{Bhattacharyya:2007vjd,Bhattacharyya:2008mz,Bagchi:2012cy,Penna:2017vms,Ciambelli:2018xat,Ciambelli:2018wre,Campoleoni:2018ltl,Ciambelli:2020eba,Ciambelli:2020ftk} for a fluid/gravity perspective.

In the second proposal, the holographic dual of gravity in $4d$ asymptotically flat spacetime is a two-dimensional ``celestial conformal field theory'' (CCFT) living on the conformal sphere at infinity. The celestial holography program is rooted on the observation that gravitational $S$-matrix elements written in a boost eigenstate basis take the form of conformal correlation functions \cite{deBoer:2003vf,Pasterski:2016qvg,Pasterski:2017kqt}. Quantum field theory soft theorems can be elegantly encoded in CCFT in terms of celestial currents associated with asymptotic symmetry generators, see \textit{e.g.} \cite{Strominger:2017zoo,Donnay:2018neh,Fotopoulos:2019tpe,Pate:2019mfs,Fan:2019emx,Donnay:2020guq,Guevara:2021abz,Strominger:2021mtt,Mago:2021wje} and \cite{Pasterski:2021raf} for more references. The obvious advantage of the celestial paradigm is that it provides a framework where one can readily make use of the plethora of powerful CFT techniques.

The main goals of this paper are $(i)$ to deepen the first picture by providing a precise proposal of a holographic description which properly takes into account leaks of gravitational radiation through the boundary and $(ii)$ to initiate a dialog between these two different approaches to flat space holography.

\begin{figure}[ht!]
\vspace{3pt}
\includegraphics[width=0.45\textwidth]{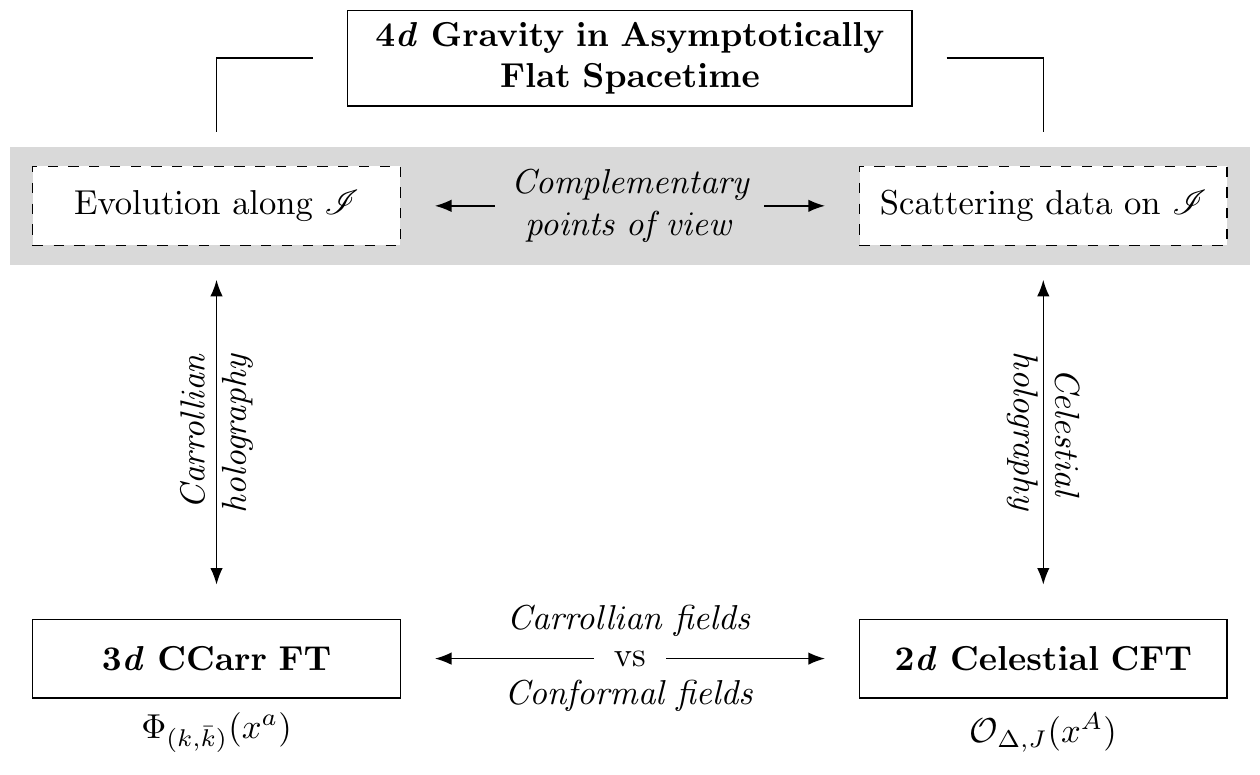}
\caption{\textit{Carrollian approach \emph{vs} Celestial approach to flat space holography}.}
\label{fig:trinity}
\end{figure}

The paper is organized as follows. In section \ref{sec:Asymptotically flat spacetimes}, after a brief review of $4d$ asymptotically flat spacetimes, we argue that the coexistence of two proposals for flat space holography relies on the two complementary descriptions of the spacetime boundary. In section \ref{sec:Sourced conformal Carrollian field theory}, we propose that the dual field theory in the first picture is a conformal Carrollian field theory  (CCarrFT) coupled with some external sources encoding the radiation leaking through the conformal boundary. In section \ref{sec:Relation with celestial holography}, we show that the Ward identities of the sourced CCarrFT reproduce those of the CCFT after taking the appropriate integral transform with respect to the advanced/retarded time. This suggests a rich set of interplays, as depicted in Figure \ref{fig:trinity}. In section \ref{sec:Discussion}, we summarize the results and discuss some implications of this work. 

\section{Asymptotically flat spacetimes}
\label{sec:Asymptotically flat spacetimes}
In this section, we review the analysis of four-dimensional asymptotically flat spacetimes at null infinity, denoted $\mathscr{I}^+$, and its relation with Carrollian geometry.

\subsection{Solution space}

In (retarded) Bondi coordinates $(u, r, x^A)$, $x^A = (z, \bar{z})$ \cite{Bondi:1962px,Sachs:1962wk,Sachs:1962zza}, the solution space of four-dimensional asymptotically flat metrics reads as \cite{Tamburino:1966zz,Barnich:2010eb} 
\begin{align}
    &ds^2 = \left(\frac{2M}{r}+\mathcal O(r^{-2})\right) {d}u^2 - 2 \left(1+\mathcal O(r^{-2})\right) {d}u {d}r \nonumber \\
    &+ \left(r^2 \mathring{q}_{AB} + r\, C_{AB} + \mathcal{O}(r^{-1})\right)dx^Adx^B \label{Bondi gauge metric} \\
    &+ \left(\frac{1}{2}\partial_B C^B_A + \frac{2}{3r}(N_A + \frac{1}{4}C_A^B\partial_C C^C_B) + \mathcal O(r^{-2}) \right)dudx^A \, , \nonumber
\end{align}
where the asymptotic shear $C_{AB}(u,z,\bar z)$ is a $2$-dimensional symmetric trace-free tensor.
For simplicity, we chose the transverse boundary metric to be the flat metric, namely $\mathring{q}_{AB} dx^A dx^B = 2 dz d\bar{z}$. The Bondi mass $M(u,z,\bar z)$ and angular momentum aspects $N_A (u,z,\bar z)$ in \eqref{Bondi gauge metric} satisfy the time evolution/constraint equations
\begin{equation}
\begin{split}
\partial_u M &= - \frac{1}{8} N_{AB} N^{AB} + \frac{1}{4} \partial_A\partial_B N^{AB} \, , \\ 
\partial_u N_A &= \partial_A M + \frac{1}{16} \partial_A (N_{BC} C^{BC}) - \frac{1}{4} N^{BC} \partial_A C_{BC} \\
&\quad  -\frac{1}{4} \partial_B (C^{BC} N_{AC} - N^{BC} C_{AC}) \\
&\quad - \frac{1}{4} \partial_B \partial^B \partial^C C_{AC}+ \frac{1}{4} \partial_B \partial_A \partial_C C^{BC} \, ,
\end{split}\label{EOM1} 
\end{equation} 
with $N_{AB} = \partial_u C_{AB}$ the Bondi news tensor encoding the gravitational radiation. 

\subsection{BMS and conformal Carrollian symmetries}

The diffeomorphisms preserving the solution space displayed above are generated by vectors fields $\xi = \xi^u \partial_u + \xi^z \partial + \xi^{\bar{z}} \bar{\partial}+ \xi^r \partial_r$ whose leading order components read as
\begin{equation}
\begin{split}
    &\xi^u =\mathcal{T} + u \alpha \, , \quad \alpha = \frac{1}{2}(\partial \mathcal{Y} +  \bar\partial \bar{\mathcal{Y}}) \, , \\
    &\xi^z = \mathcal{Y} + \mathcal{O}(r^{-1})\, , \qquad \xi^{\bar{z}} = \bar{\mathcal{Y}} + \mathcal{O}(r^{-1}) \, , \\ 
    &\xi^r = - r \alpha + \mathcal{O}(r^0) \, ,
    \end{split}
    \label{AKV Bondi}
\end{equation} where $\mathcal{T}= \mathcal{T}(z, \bar{z})$ is the supertranslation parameter and $\mathcal{Y} = \mathcal{Y}(z)$, $\bar{\mathcal{Y}} = \bar{\mathcal{Y}}(\bar{z})$ are the superrotation parameters satisfying the conformal Killing equation. Using a modified Lie bracket \cite{Barnich:2010eb}, the asymptotic Killing vectors \eqref{AKV Bondi} satisfy the (extended) BMS algebra.

The infinitesimal transformation of the asymptotic shear $C_{AB}$ under BMS symmetries can be split into hard and soft pieces $\delta_\xi C_{zz} = \delta^H_\xi C_{zz} + \delta^S_\xi C_{zz}$ which are respectively homogeneous and inhomogeneous in $C_{zz}$ \cite{Strominger:2013jfa,He:2014laa}. This reads explicitly as
\begin{align}
    &\delta^H_\xi C_{zz} = \left[\left(\mathcal{T} + u\alpha\right) \partial_u  + \mathcal Y \partial + \bar{\mathcal Y}\bar{\partial} +   \frac{3}{2} \partial \mathcal Y - \frac{1}{2} \bar{\partial} \bar{\mathcal Y}\right] C_{zz} \, , \nonumber\\
    &\delta^S_\xi C_{zz} = -2 \partial^2 \mathcal{T} - u\, \partial^3 \mathcal Y \, , \label{transfo CAB}
\end{align} together with the complex conjugate relations for $C_{\bar{z}\bar{z}}$.

From a geometric perspective, BMS symmetries are the conformal symmetries of a Carrollian structure on $\mathscr{I}^+$ with coordinates $x^a = (u, z, \bar{z})$~\cite{Duval:2014uva,Duval:2014lpa} (see also \cite{1977asst.conf....1G,Henneaux:1979vn, Ashtekar:2014zsa,Ciambelli:2019lap,Figueroa-OFarrill:2019sex, Herfray:2020rvq,Herfray:2021xyp, Herfray:2021qmp,Henneaux:2021yzg}). This Carrollian structure is given by a degenerate metric $q_{ab}$ and a vector field $n^a$ in the kernel of $q_{ab}$,  \textit{i.e.} $q_{ab} n^b = 0$. From \eqref{Bondi gauge metric}, it reads explicitly  $q_{ab} dx^a dx^b = 0 du^2 + 2 d z d\bar{z}$ and $n^a \partial_a = \partial_u$. The conformal Carrollian symmetries are generated by vector fields $\bar\xi = \bar \xi^a \partial_a$ on $\mathscr{I}^+$ satisfying
\begin{equation}
    \mathcal{L}_{\bar{\xi}} q_{ab} = 2 \alpha q_{ab}\, , \qquad \mathcal{L}_{\bar\xi} n^a = - \alpha n^a \, .
    \label{conformal Carroll symmetries}
\end{equation} The solution $\bar{\xi}$ of \eqref{conformal Carroll symmetries} is precisely given by the restriction to $\mathscr{I}^+$ of the asymptotic Killing vectors \eqref{AKV Bondi},  \textit{i.e.} 
\begin{equation}
    \bar{\xi} = (\mathcal{T} + u \alpha )\partial_u + \mathcal{Y} \partial + \bar{\mathcal{Y}} \bar{\partial} \, .
   \label{conformal Carroll symmetries here}
\end{equation} The standard Lie bracket on $\mathscr{I}^+$ of these vector fields reproduces the BMS algebra.

\subsection{BMS charges}

 At each cut $\mathcal{S} \equiv \{u = \text{constant}\}$ of $\mathscr{I}^+$, the charges associated with the BMS symmetries \eqref{AKV Bondi} are given by \cite{Strominger:2013jfa,He:2014laa, Kapec:2014opa,Hawking:2016sgy,Compere:2018ylh,Campiglia:2020qvc,Compere:2020lrt,Fiorucci:2021pha,Donnay:2021wrk}
\begin{align}
&Q_\xi = \kappa \int_{\mathcal{S}} dz d\bar{z}\, \left(4 \mathcal{T} M + 2 \mathcal Y^A \bar N_A\right), \nonumber \\
&\bar N_A = N_A - u\, \partial_A M + \frac{1}{4}C_A^B \partial_C C_B^C +\frac{3}{32}\partial_A (C_B^CC^B_C) \nonumber \\
&\qquad + \frac{u}{4}  \partial^B \partial_B \partial_C C_A^C - \frac{u}{4} \partial^B \partial_A \partial_C C_B^C 
 \label{grav charge}
\end{align}
where $\mathcal Y^A = (\mathcal{Y}, \bar{\mathcal{Y}})$ and $\kappa = (16\pi G)^{-1}$. These charges differ from the prescriptions considered \textit{e.g.} in \cite{Dray:1984rfa,Wald:1999wa,Barnich:2011mi,Flanagan:2015pxa}; we refer to \cite{Compere:2019gft} for a recent discussion on the relation between the various proposals. Using the appropriate bracket \cite{Barnich:2011mi} (see also \cite{Troessaert:2015nia, Compere:2018ylh,Freidel:2021fxf,Freidel:2021cjp, Wieland:2021eth, Freidel:2021dxw,Chandrasekaran:2021vyu}), these charges satisfy an algebra from which flux-balance laws can be deduced. The charges \eqref{grav charge} are conserved in absence of outgoing radiation, namely when $N_{AB} = 0$.

Notice that the analysis of asymptotically flat spacetimes can be performed at $\mathscr{I}^-$ in an analogous manner by working in advanced Bondi coordinates $(v,r,x^A)$.

\subsection{Holographic nature of null infinity}

From the analysis of $4d$ asymptotically flat spacetimes, two complementary pictures emerge, which correspond to the two possible roads to flat space holography, as discussed in the introduction (see Figure \ref{fig:trinity}):

\begin{itemize}
    \item[$\rhd$] In the first picture, $\mathscr{I}^+$ (resp. $\mathscr{I}^-$) is seen as a boundary along which there is retarded (resp. advanced) time evolution. This is suited to describe the dynamics of the system through flux-balance equations such as the famous Bondi mass loss formula originally derived in~\cite{Bondi:1962px,Sachs:1962wk,Sachs:1962zza}. In other terms, equations \eqref{EOM1} are interpreted as evolution equations, suggesting a $4d$ bulk / $3d$ boundary holographic correspondence, where the dual field theory lives at null infinity and obeys Carrollian physics. Since the charges are generically not conserved due to the outgoing (resp. ingoing) radiation going through null infinity, it is tempting to think of the dual theory as coupled to some external sources responsible for the dissipation. We elaborate on this proposal in section \ref{sec:Sourced conformal Carrollian field theory}. 
    
    \item[$\rhd$] In the second picture, $\mathscr{I}^+$ (resp. $\mathscr{I}^-$) is seen as a portion of a Cauchy surface in the asymptotic future (resp. past). This point of view is well adapted to describe the scattering problem in asymptotically flat spacetime between $\mathscr{I}^-$ and $\mathscr{I}^+$ and provides information about the state of the system at early and late times. The equivalence between BMS Ward identities and soft theorems was established in this picture \cite{Strominger:2013jfa,He:2014laa, Kapec:2014opa}. Equations \eqref{EOM1} are now seen as constraint equations in the Hamiltonian framework. Scattering amplitudes in the bulk can be rewritten as correlation functions on the celestial sphere obeying some Ward identities encoding the information on soft theorems. This suggests a $4d$ bulk / $2d$ boundary correspondence, with a $2d$ CCFT as holographic dual. We come back on this proposal in section \ref{sec:Relation with celestial holography}, where we relate it to the Carrollian framework. 
\end{itemize}

\section{Sourced conformal Carrollian field theory}
\label{sec:Sourced conformal Carrollian field theory}

In this section, we write the Ward identities for a sourced quantum field theory. We then specify this result for a sourced CCarrFT and argue that it holographically encodes the asymptotic dynamics of gravity.

\subsection{Sourced Ward identities}

Let us start with a theory of fields $\Phi^i$ on a $n$-dimensional manifold $\mathscr M$ with coordinates $x^a$ and admitting a well-defined variational principle. We assume that the theory exhibits some global symmetries $\delta_K \Phi^i = K^i[\Phi]$ with associated conserved Noether currents $j^a_K$. If one couples the theory to external sources $\sigma$ which are non-dynamical fields, the infinitesimal transformations $\delta_K \Phi^i = K^i[\Phi]$ are generically no longer symmetries due to the presence of the sources (see \textit{e.g.} \cite{Troessaert:2015nia,Wieland:2020gno,ToAppear}). This translates into the fact that the Noether currents are no longer conserved, namely
\begin{equation}
    \partial_a j^a_K(x) = F_K(x) \, ,
    \label{generic flux-balance}
\end{equation} where the local flux $F_K(x)$, which generically depends on $\Phi$ and $\sigma$, vanishes when $\sigma = 0$. At the quantum level, taking the presence of external sources into account in the standard derivation of the Ward identities\footnote{See \textit{e.g.} \cite{DiFrancesco:1997nk} for the standard derivation of Ward identities. The consideration of external sources in this derivation will be detailed in an upcoming work.}, we obtain
\begin{equation}
\begin{aligned}
&\partial_a \langle j^a_K(x) X \rangle + \frac{\hbar}{i}\sum_{i=1}^N \delta^{(n)}(x-x_i) \, \delta_{K^i} \braket{X} \\
&\qquad =  \braket{ F_K(x) X }\,,
\end{aligned} \label{sourced ward identities}
\end{equation} where $X \equiv \Phi^{i_1}(x_1)\dots\Phi^{i_N}(x_N)$ denotes the collection of quantized fields and $\delta_{K^i} \braket{X} \equiv \braket{\Phi^{i_1}(x_1)\dots K^{i}[  \Phi (x_i)]\dots \Phi^{i_N}(x_N)}$. This generalizes the local version of the infinitesimal Ward identities in presence of external sources. In particular, with no field insertion in the correlators, it implies
\begin{equation}
    \partial_a \langle j^a_K(x) \rangle = \braket{F_K(x)}\,,
\end{equation} which reproduces the classical flux-balance equation \eqref{generic flux-balance}. Integrating \eqref{sourced ward identities} over the manifold $\mathscr{M}$ with boundary $\partial \mathscr{M}$, we get the integrated version of the infinitesimal Ward identities with external sources:
\begin{equation}
    \begin{split}
        &\sum_{i=1}^N  \delta_{K^i} \braket{X}  = \frac{i}{\hbar} \Big\langle \left( \int_\mathscr{M} \bm F_K - \int_{\partial \mathscr M} \bm j_K \right) X \Big\rangle\,,
    \end{split} \label{integrated sourced Ward identity}
\end{equation} where bold letters denote the forms associated with the objects defined above, \textit{e.g.} $\bm F_K = F_K\, (d^n x)$ and $\bm j_K = j_K^a \,(d^{n-1}x)_a$. The standard result of the invariance of the correlators under symmetry transformations is recovered by turning off the sources and assuming that the Noether currents vanish at the boundary.

\subsection{Application to conformal Carrollian field theory}

We now apply these general results to the case of a $3d$ CCarrFT coupled to some external sources. 

BMS symmetries are to $3d$ CCarrFT  what Virasoro symmetries are to $2d$ CFT, while the $4d$ Poincar\'e group is on the same footing as the M\"obius group $SL(2,\mathbb C)$. ``Global" subgroups inside BMS and Virasoro groups are not unique and correspond to Poincar\'e and M\"obius subgroups, respectively. Accordingly, conformal Carrollian primary fields \cite{Ciambelli:2018xat,Ciambelli:2018wre,Ciambelli:2019lap} will be taken to transform infinitesimally as
\begin{equation}
    \begin{split}
        &\delta_{\bar{\xi}} \Phi_{(k,\bar{k})}
        = [(\mathcal{T} + u \alpha )\partial_u + \mathcal Y \partial + \bar{\mathcal Y}\bar{\partial} \\
        &\qquad\qquad\quad +   k\, \partial \mathcal 
        Y + \bar{  k}\, \bar{\partial} \bar{\mathcal Y}] \Phi_{(  k,\bar{  k})}
    \end{split}\label{Carrollian tensor def}
\end{equation}
under full conformal Carroll symmetries \eqref{conformal Carroll symmetries here} while quasi-primary fields are only required to transform properly under the global subgroup. Here the Carrollian weights $(k,\bar{k})$ are some integers or half-integers.

Noether currents associated with the conformal Carrollian symmetries \eqref{conformal Carroll symmetries here} are taken to be of the form
\begin{equation}
    j^a_{\bar \xi} = {\mathcal{C}^a}_{b} \bar{\xi}^b 
    \label{Carrollian Noether currents}
\end{equation} where ${\mathcal{C}^a}_b$ is the analogue of the stress-energy tensor for a CCarrFT encoding the Carrollian momenta \cite{Hartong:2015usd,deBoer:2017ing,Ciambelli:2018xat, Ciambelli:2018wre, Ciambelli:2018ojf, Donnay:2019jiz,deBoer:2021jej,Chandrasekaran:2021hxc,Freidel:2021qpz,Freidel:2021dfs} as follows:
\begin{align}
    {\mathcal{C}^a}_{b} = \left[
        \begin{array}{cc}
            \mathcal{M} &\mathcal{N}_B \\
            \mathcal{B}^A &{\mathcal{A}^A}_{B}
        \end{array} 
    \right].
\end{align} The Noether currents \eqref{Carrollian Noether currents} associated with the global subalgebra of the conformal Carrollian algebra verify the flux-balance law \eqref{generic flux-balance} provided ${\mathcal{C}^a}_b$ satisfies
\begin{equation}
    \begin{array}{rcl}
        \text{Translations}\, (\partial_a) & \Rightarrow & \partial_a {\mathcal{C}^a}_{b} = F_b \, , \\
        \text{Rotation}\,  (-z\partial+\bar z\bar\partial) & \Rightarrow & {\mathcal{C}^z}_{z} = {\mathcal{C}^{\bar{z}}}_{\bar{z}} \, , \\
        \text{Boosts}\, (x^A\partial_u) & \Rightarrow & {\mathcal{C}^A}_u = 0\, ,\\
        \text{Dilatation}\,  (x^a\partial_a) & \Rightarrow & {\mathcal{C}^a}_a = 0\, ,
    \end{array}
\label{classical constraints on the C}
\end{equation}
where we assumed that the flux is linear in the parameters $\bar{\xi}^a$ and can be written as $F_{\bar{\xi}} = F_a \bar\xi^a$ in the right-hand side of \eqref{generic flux-balance}, which is sufficient for the purpose of this paper. Notice that the term ${\mathcal C^a}_b\partial_a\bar \xi^b$ does not contribute to the left-hand side of \eqref{generic flux-balance} as a consequence of \eqref{conformal Carroll symmetries here} and \eqref{classical constraints on the C}. The Carrollian special conformal transformations $K_0 = -2z\bar z\partial_u$, $K_1 = 2 u \bar z\partial_u + 2 \bar z^2\bar\partial$ and $K_2 = 2 u z \partial_u + 2 z^2\partial$ do not impose further constraints. Furthermore, the above global conformal Carrollian symmetries are enough to completely constrain ${\mathcal{C}^a}_b$, \textit{i.e.} \eqref{generic flux-balance} is automatically satisfied by the supertranslation (and superrotation) currents provided \eqref{classical constraints on the C} holds. In terms of the Carrollian momenta, the constraints \eqref{classical constraints on the C} imply
\begin{equation}
\begin{array}{rl}
    \partial_u \mathcal{\mathcal{M}} = F_u\, , &\quad \mathcal{B}^z =0\, , \\
    \partial_u \mathcal{N}_z - \frac{1}{2}\partial \mathcal{M} + \bar\partial {\mathcal{A}^{\bar{z}}}_z =  F_z \, , &\quad 2 {\mathcal{A}^z}_z +  \mathcal{M} = 0\,,
\end{array}
\label{flux balance carrollian momenta}
\end{equation} 
together with the complex conjugate relations.

At the quantum level, one can write the infinitesimal Ward identities \eqref{sourced ward identities} for the specific case of a CCarrFT. Assuming that the operators inserted in the correlators are (quasi-)conformal Carrollian primary fields, we obtain
\begin{align}
        &\partial_u \braket{\mathcal M\, X} + \frac{\hbar}{i}\sum_i \delta^{(3)}(x-x_i)\partial _{u_i} \braket{X} = \braket{F_u\, X}\,, \nonumber \\
        &\partial_u \braket{\mathcal N_z\, X} - \frac{1}{2}\partial\braket{\mathcal M\, X} + \bar\partial\braket{{\mathcal A^{\bar z}}_z\, X} \nonumber \\
        &+ \frac{\hbar}{i}\sum_i \left[ \delta^{(3)}(x-x_i) \partial_i \braket{X} - \partial \left(\delta^{(3)}(x-x_i)\, k_i \braket{X}\right)\right] \nonumber \\
        &=\braket{F_z\, X} \,,\qquad \braket{\mathcal{B}^z\,X} = 0 \,, \nonumber \\
        &\braket{({\mathcal A^z}_z+ \frac{1}{2}\mathcal M)X} + \frac{\hbar}{i}\sum_i \delta^{(3)}(x-x_i)\,k_i\,\braket{X} = 0\, ,  \label{Ward identities Carrollian momenta}
\end{align}
together with the complex conjugate relations. With no field insertion in the correlators, the expectation value of the operators reproduce the classical relations \eqref{flux balance carrollian momenta}. At any stage of this analysis, one can turn off the sources by setting $F_a =0$ to obtain the Ward identities of an honest $3d$ CCarrFT.

\subsection{Holographic correspondence}

We now argue that a quantum CCarrFT coupled with external sources would be a right candidate to describe holographically gravity in $4d$ asymptotically flat spacetimes reviewed in section \ref{sec:Asymptotically flat spacetimes} (few explicit examples of quantum CCarrFT are known, see \textit{e.g.} \cite{Isberg:1993av,Mason:2013sva,Bagchi:2015nca,Hao:2021urq,Chen:2021xkw}). We propose the following correspondence between Carrollian momenta and gravitational data at $\mathscr{I}^+$:
\begin{equation}
\begin{split}
        &\langle \mathcal{M} \rangle = 4\,\kappa\, M \, , \\
        &\langle \mathcal{N}_A \rangle =  2\,\kappa\, \Big( N_A + \frac{1}{4}C_A^B \partial_C C_B^C +\frac{3}{32}\partial_A (C_B^CC^B_C) \\
        &\qquad\qquad + \frac{u}{4}  \partial^B \partial_B \partial_C C_A^C - \frac{u}{4} \partial^B \partial_A \partial_C C_B^C \Big) \, , \\
        &\langle {\mathcal{A}^A}_{B} \rangle + \frac{1}{2}{\delta^A}_B \langle \mathcal M \rangle = 0 \, . 
\end{split}
\label{holographic correspondence}
\end{equation} 
The factors are fixed by demanding that the gravitational charges \eqref{grav charge} correspond to the Noether currents \eqref{Carrollian Noether currents} of the CCarrFT integrated on a section $u=\text{constant}$. The correspondence \eqref{holographic correspondence} is reminiscent of the AdS/CFT dictionary where the holographic stress-energy tensor of the CFT is identified with some subleading order in the expansion of the bulk metric \cite{Balasubramanian:1999re,deHaro:2000vlm}. It would be interesting to push this analogy further and see if the Carrollian momenta can be obtained by varying a bulk partition function with respect to the
boundary sources at null infinity (see \textit{e.g.} \cite{Detournay:2014fva,Bagchi:2015wna} for discussions along these lines in $3d$ gravity). 

The Bondi news tensor $N_{AB}$ is a free datum at $\mathscr{I}^+$ that encodes the outgoing gravitational radiation. It is responsible for the non-conservation of the BMS charges \eqref{grav charge} at null infinity. It is therefore suggestive to holographically identify the external sources $\sigma_{AB}$ at the boundary with the Bondi news tensor $N_{AB}$ as
\begin{equation}
    \sigma_{AB} = N_{AB} \,.
\end{equation} The external sources are responsible for the dissipation in the CCarrFT through the fluxes 
\begin{align}
        F_u &= -\kappa\Big[ \sigma^{zz}\sigma_{zz} - 2 (\bar\partial^2 \sigma_{zz}+\partial^2 \sigma_{\bar z\bar z}) \Big]\,, \label{flux expressions sources} \\
        F_z &= \frac{\kappa}{2}\Big[ \partial (\sigma^{zz}\Phi_{zz}) + 2 \Phi_{zz}\partial \sigma^{zz} + u \partial(\bar\partial^2 \sigma_{zz}-\partial^2 \sigma_{\bar{z}\bar{z}}) \Big]\, .\nonumber
\end{align} 
Here $\Phi_{zz}$ denotes the operator associated with a perturbation in the asymptotic shear: $\langle \Phi_{AB} \rangle = C_{AB}$. Comparing \eqref{transfo CAB} with \eqref{Carrollian tensor def}, one deduces that $\Phi_{zz}$ is a quasi-conformal Carrollian primary fields of weights $(\frac{3}{2}, - \frac{1}{2})$. It constitutes a particular example of a correlator insertion. Notice that, from the boundary perspective, the momentum $\Pi_{AB} = \partial_u  \Phi_{AB}$ conjugated to $\Phi_{AB}$ should be distinguished from the sources $\sigma_{AB}$. They are identified only through VEV as $\langle \Pi_{AB} \rangle = \sigma_{AB}$. 

Taking the identifications \eqref{holographic correspondence} and \eqref{flux expressions sources} into account, one can then check explicitly that the time evolution equations in the sourced Ward identities \eqref{Ward identities Carrollian momenta} reproduce the gravitational retarded time evolution equations \eqref{EOM1} when there is no insertion in the correlators.

Let us emphasize that a similar identification can be performed with the solution space in advanced Bondi coordinates $(v, r, x^A)$ at $\mathscr{I}^-$. It is therefore natural to assume that the dual sourced CCarrFT is living on $\hat{\mathscr I} = \mathscr{I}^- \sqcup \mathscr{I}^+$, where the two manifolds $\mathscr{I}^-$ and $\mathscr{I}^+$ are glued together by identifying antipodally $\mathscr{I}^+_{-}$ with $\mathscr{I}^-_{+}$. Geometrically, the gluing 2-sphere on $\hat{\mathscr I}$ is distinguished by the vanishing of the vector field $n^a$ defining the Carrollian structure. Indeed, $\mathscr{I}^+_-$ and $\mathscr{I}^-_+$ are stable under supertranslation. The gluing is consistent with the antipodal matching conditions proposed in \cite{Strominger:2013jfa,He:2014laa, Kapec:2014opa} and confirmed in \cite{Troessaert:2017jcm,Compere:2017knf,Henneaux:2018cst, Henneaux:2018hdj, Prabhu:2019fsp,Prabhu:2021cgk} by an analysis at spacelike infinity. In particular, the Carrollian data are identified with the solution space of the retarded (resp. advanced) Bondi gauge at $\mathscr{I}^+$ (resp. $\mathscr{I}^-$), with a continuous interpolation in the gluing region thanks to the antipodal matching. The conformal Carrollian symmetries act on the whole $\hat{\mathscr I}$ and correspond to the diagonal BMS symmetries identified in \cite{Strominger:2013jfa}. They are generated by \eqref{conformal Carroll symmetries} on $\mathscr{I}^+$ and the analogue antipodally matched symmetries on $\mathscr{I}^-$.

\section{Relation with celestial holography}
\label{sec:Relation with celestial holography}

In this section, we show that the Ward identities of the sourced CCarrFT reproduce the BMS Ward identities of the celestial CFT after performing the right integral transformations.
This constitutes a central argument to relate the two approaches of flat space holography, see Figure \ref{fig:trinity}.

Specifying the integrated version of the sourced Ward identities \eqref{integrated sourced Ward identity} to the conformal Carrollian symmetries of the theory living on $\hat{\mathscr{I}}$ suggested in the previous section, we obtain 
\begin{equation}
   \begin{split}
        & \delta_{\bar{\xi}} \langle X  \rangle  = \frac{i}{\hbar} \Big\langle \left( \int_{\hat{\mathscr{I}}}  {\bm F}_{\bar{\xi}} - \int_{\mathscr{I}^+_+}  {\bm j}_{\bar{\xi}} + \int_{\mathscr{I}^-_-}  {\bm j}_{\bar{\xi}} \right) \, X \Big\rangle \, ,
    \end{split} \label{stage Ward}
\end{equation} where $X  \equiv  \Phi^{out}_{(k_1, \bar{k}_1)}(x_1)\dots \Phi^{out}_{(k_m, \bar{k}_m)}(x_m) \Phi^{in}_{(k_1,\bar{k}_1)}(x_1) \dots $ $ \Phi^{in}_{(k_n,\bar{k}_n)}(x_n)$, $\Phi^{out}_{(k_i,\bar{k}_i)}(x_i)$ and $\Phi^{in}_{(k_j,\bar{k}_j)}(x_j)$ denoting insertions at $\mathscr{I}^+$ and at $\mathscr{I}^-$, respectively. To simplify the discussion, let us assume that we are describing a scattering of massless particles, so that the current $\bm j_{\bar{\xi}}$ vanishes at $\mathscr{I}^+_+$ and $\mathscr{I}^-_-$. In addition, we require that the integrated flux on $\mathscr{I}^+$ is equal to minus the integrated flux on $\mathscr{I}^-$ at the level of the operators, namely
\begin{equation}
    \int_{\mathscr{I}^-}  {\bm F}_{\bar{\xi}}= -  \int_{\mathscr{I}^+}  {\bm F}_{\bar{\xi}} \, .
\end{equation} This constraint on the sources is compatible with the classical bulk requirement that the integrated ingoing flux is equal to the integrated outgoing flux for the massless scattering considered. Taking these assumptions into account, the integrated Ward identities \eqref{stage Ward} imply
\begin{equation}
    \delta_{\bar{\xi}} \langle X  \rangle  = 0\,. \label{carroll invariance}
\end{equation} This is the statement that the correlators are conformal Carroll invariant. The consequence of this relation have been studied  \textit{e.g.} in \cite{Chen:2021xkw,Bagchi:2009ca,Bagchi:2017cpu}.

To relate \eqref{carroll invariance} with the CCFT Ward identities, we use similar technical steps than those advocated in \cite{Strominger:2013jfa,He:2014laa,Kapec:2014opa} to relate BMS Ward identities and soft graviton theorems in the bulk. Let us first specify \eqref{carroll invariance} for the supertranslation symmetries to recover the corresponding Ward identities of the CCFT. We split the variation into hard and soft parts $\delta_{\mathcal{T}} \langle X  \rangle = \delta^H_{\mathcal{T}} \langle X  \rangle +  \delta^S_{\mathcal{T}} \langle X  \rangle$ and rewrite the soft part of the transformation as the soft charge insertion by using the quantum commutator
\begin{equation}
\begin{split}
&\left[\Pi_{zz}(u, z, \bar{z}), \Phi_{\bar{w}\bar{w}}(u',w, \bar{w})\right] = \frac{i\hbar}{\kappa}  \, \delta (u- u')\, \delta^{(2)} (z-{w})\,,
\end{split}
    \label{commutation PiPhi}
\end{equation} which ensures the compatibility with the Poisson bracket on the radiative phase space \cite{Ashtekar:1981sf,Ashtekar:1981bq}. Then we specify the relation for $\mathcal{T}(z, \bar{z}) = \frac{1}{z-w}$ and introduce the supertranslation current $P(z, \bar{z})$ \cite{Strominger:2013jfa} through
\begin{equation}
    P (z, \bar{z})= \frac{1}{4 G}\left(\int du + \int dv  \right) \bar{\partial} \Pi_{zz}\,.
\end{equation} 
Furthermore, we obtain the CCFT operators $\mathcal{O}^{out}_{\Delta,J}(z,\bar{z})$ and $\mathcal{O}^{in}_{\Delta,J}(z,\bar{z})$ of conformal dimension $\Delta$ and spin $J$ from the conformal Carrollian operators $\Phi^{out}_{(k,\bar{k})}(u,z,\bar{z})$ and $\Phi^{in}_{(k,\bar{k})}(v,z,\bar{z})$ of \eqref{Carrollian tensor def} through the integral transform
\begin{equation}
\begin{split}
     \mathcal{O}^{out}_{\Delta,J} (z, \bar{z})  &= i^{\Delta}\Gamma[\Delta] \int^{+\infty}_{-\infty} du\, u^{-\Delta}\, \Phi^{out}_{(k,\bar k)}(u,z,\bar z)\, , \\
     \mathcal{O}^{in}_{\Delta,J} (z, \bar{z})  &= i^{\Delta}\Gamma[\Delta] \int^{+\infty}_{-\infty} dv\, v^{-\Delta}\, \Phi^{in}_{(k,\bar k)}(v,z,\bar z)\,.
     \label{integral transformss}
\end{split}
\end{equation}
The above integral is the composition of a Fourier transform (from position to momentum space) and a Mellin transform which maps energy to boost eigenstates \cite{Pasterski:2016qvg,Pasterski:2017kqt}. It trades the time dependence of the Carrollian operators for the conformal dimension of the CCFT operators. Importantly, Carrollian weights are related to the $2d$ spin via
\begin{equation}
    k=\frac{1}{2}(1+J)\,,\qquad \bar{k}=\frac{1}{2}(1-J)\,,
\end{equation}
which can be seen to be consistent with the radiative falloffs in the conformal compactification. Taking into account the aforementioned steps, we recover the CCFT Ward identity for supertranslations \cite{Strominger:2013jfa,He:2014laa,Donnay:2018neh} ($N=m+n$)
\begin{equation}
    \begin{split}
& \Big\langle P(z,\bar{z}) \prod_{i=1}^N \mathcal{O}_{\Delta_i,J_i} (z_i, \bar{z}_i) \Big\rangle \\
&+ \hbar \sum_{q=1}^{N} \frac{1}{z-z_q}  \Big\langle \dots \mathcal{O}_{\Delta_{q} +1,J_q} (z_q, \bar{z}_q)  \dots \Big\rangle = 0 \,,
\end{split} 
\end{equation} 
which is the celestial encoding of the leading soft graviton theorem.

Now, one can specify \eqref{carroll invariance} for superrotations and follow the same series of steps. Choosing $\mathcal{Y}(z) = \frac{1}{z-w}$ and defining the $2d$ stress-tensor as
\begin{equation}
T (z) = -\frac{i}{8\pi G}\int  \frac{dw d\bar{w}}{z-w} \left(\int du \, u + \int dv \, v \right) \partial^3\Pi_{\bar{w}\bar{w}}\, ,
\end{equation}
we recover the $2d$ CFT Ward identities \cite{Kapec:2016jld,Cheung:2016iub,Fotopoulos:2019tpe,Fotopoulos:2019vac} after performing \eqref{integral transformss}:
\begin{align}
& \Big\langle T(z) \prod_{i=1}^N \mathcal{O}_{\Delta_i,J_i} (z_i, \bar{z}_i)  \Big\rangle \\
&+{\hbar}\sum_{q=1}^N \Big[ \frac{\partial_q}{z-z_q}  + \frac{h_q}{(z-z_q)^2} \Big] \, \Big\langle \prod_{i=1}^N \mathcal{O}_{\Delta_i,J_i} (z_i, \bar{z}_i) \Big\rangle=0 \,,\nonumber
\end{align} 
where $h_q=\tfrac{1}{2}(\Delta_q+J_q)$ (similar results hold for the anti-holomorphic stress tensor). This result encodes the information on the subleading soft graviton theorem in the bulk.

\section{Discussion}
\label{sec:Discussion}

Here, we have argued that the null nature of $\mathscr{I}$ leads to two complementary roads to flat space holography (see Figure \ref{fig:trinity}): Carrollian \textit{versus} Celestial. We have presented a holographic description in the first picture in terms of a codimension-one CCarrFT coupled with external sources encoding gravitational radiation at null infinity. The Ward identities have been shown to reproduce those of the celestial CFT, by relating celestial primary operators to conformal Carrollian fields living at $\mathscr{I}$. This provides a bridge between two different approaches to the ambitious program of finding a holographic description of quantum gravity for realistic spacetimes.

This work raises new questions for the future. For instance, it would be interesting $(i)$ to deduce the low-point correlation functions in the CCarrFT from the Ward identities and match them with those of the CCFT, $(ii)$ to understand if the sourced CCarrFT can be obtained in the flat limit of the AdS/CFT correspondence (building on the works \cite{Compere:2019bua,Compere:2020lrt,Fiorucci:2020xto}), $(iii)$ to write a concrete proposal for the sourced CCarrFT reproducing the features described in this work. We let this research program for future investigations.

\begin{acknowledgments}
We would like to thank Glenn Barnich, Geoffrey Comp\`ere, Florian Ecker, Gaston Giribet, Daniel Grumiller and Marios Petropoulos for discussions. LD, AF and RR are supported by the Austrian Science Fund (FWF) START project Y 1447-N. AF and RR also acknowledge support from the FWF, project P 32581-N. YH has received funding from the European Research Council (ERC) under the European Union’s Horizon 2020 research and innovation programme (grant agreement No 101002551). The authors also acknowledge support of the Erwin Schrödinger Institute (ESI) in Vienna where part of this work was conducted during the thematic program ``Geometry for Higher Spin Gravity$\_$FCC$\_$2021''.
\end{acknowledgments}

\bibliographystyle{style}

\providecommand{\href}[2]{#2}\begingroup\raggedright\endgroup

\end{document}